\documentstyle[prl,multicol,aps]{revtex} 

\begin{document}
\draft
\title
{Rigorous results for a hierarchy of generalized Heisenberg models}

\author
{You-Quan Li} 
\address
{Zhejiang Institute of Modern Physics,
Zhejiang University, Hangzhou 310027, China and\\
Institut f\"ur Physik, Universit\"at Augsburg, 
D-86135 Augsburg, Germany}

\date{Received: Jan 19, 2001}

\maketitle

\begin{abstract}
The Lieb-Schultz-Mattis theorem is extended to generalized
Heisenberg models related to non-exceptional Lie algebras.
It is shown that there are no energy gaps above the ground sates
for SO(4), Sp(2) and SU(4) Heisenberg models, gaps are suspected
to occur in SO(5) and SO(6) models. The nondegenerate 
ground state for these models is rigorously proven.
\end{abstract}



\begin{multicols}{2}

The study of the ground state and excitations for 
many-particle systems is of importance, being  
relevant to superconductivity as well as Mott-transitions.
Based on Marshall's rule \cite{Marshall}, Lieb, Schultz and Mattis
proved \cite{LSM} a remarkable theorem: the spin 1/2 system
with Heisenberg interaction favors an antiferromagnetic ordering,
its ground state is nondegenerate and no energy gap exists above the 
ground state in its energy spectrum. 
Haldane pointed out \cite{Haldane} by means of a mapping into 
a nonlinear $\sigma$ model that there will be a gap to the 
excited state for the system with integer spin. 
The Lieb-Schultz-Mattis (LSM) theorem was extended 
by Kolb \cite{Kolb} and by Affleck and Lieb \cite{ALieb} to arbitrary
half-odd-integer spin, demonstrating a difference between integer
spin and half-odd-integer spin, 
in the agreement with Haldane's conjecture. 
Very recently the LSM theorem was extended to the case with an
applied external field \cite{OYAffleck97a}. It was also discussed
\cite{OYAffleck97b} in a generalized single-band Hubbard model.
Actually, ref. \cite{ALieb} also made an extension to 
SU($2n$) model by placing self-conjugate 
representation on each lattice site.
However, the case of fundamental representation 
which becomes very important nowadays due to the model study on
the spin systems with orbital degeneracy\cite{LiMSZ} was not investigated.
Additionally, a nondegenerate ground state has been proposed  
\cite{ALieb,OYAffleck97a,OYAffleck97b}
but not rigorously proven for various models except in the
original SU(2) model \cite{LSM}. 

Since various symmetries, as in SO(5) model \cite{Zhang} for
high $T_c$ superconductivity and SU(4) model \cite{LiMSZ}
for orbital physics \cite{TNagaosa} have been of interest 
lately, we consider general Heisenberg-type models
related to the non-exceptional Lie algebras \cite{Gilmore}.
We explicitly study several cases by a procedure analogous
to the LSM theorem. Some rigorous results on 
SO(4), Sp(2), SO(5), SU(4) and SO(6) Heisenberg models are given.
The ground state of those models are rigorously proven to be nondegenerate.
It is shown that there are gapless excitations in SO(4), Sp(2)
and SU(4) models. For SO(5) and SO(6), however, an energy gap is 
expected to occur. 
In other words,  the SO(4), Sp(2) and SU(4) models satisfy the sufficient
condition for gapless excitation but the  SO(5) and SO(6) models 
obey the necessary condition for the existence of a gap.

We consider a generalized Heisenberg model:
\begin{equation}
{\cal H}=\sum_{\stackrel{<x,x'>}{m n}}g^{mn}H_m(x)H_n(x')
        +\sum_{\stackrel{<x,x'>}{\alpha\in\Delta}}E_\alpha(x)E_{-\alpha}(x').
\label{eq:Hamiltonian}
\end{equation}
where $<x,x'>$ stands for nearest neighbor-pairing and
$\Delta$ denotes the set of roots of some non-exceptional Lie algebra
which will be specified in our discussion later on.
$H_m(x)$ and $E_\alpha(x)$ are generators of the Lie algebra on site $x$
in a lattice. The  $\{ H_m\}$ are the generators in the corresponding
Cartan subalgebra. We  adopt the following defining relations in 
our present paper, 
\begin{eqnarray}
\bigl[ H_m, H_n\bigr]&=&0,\;\;\;\bigl[E_{\alpha_m}, E_{-\alpha_m}\bigr]=2H_m,
 \nonumber\\[1mm] 
\bigl[E_\alpha, E_\beta\bigr]&=&E_{\alpha+\beta},
 \;{\rm  if }\; \alpha+\beta \in \Delta\nonumber\\[1mm]
\bigl[H_m, E_\alpha\bigr]&=&(\alpha)_m E_\alpha,   
\label{eq:ChavelleyBasis}
\end{eqnarray}
where $(\alpha)_m:=\alpha\cdot\alpha_m$ is the $m$-th 
{\it covariant component}
of the root vector $\alpha$ in the nonorthogonal coordinates in which the 
simple roots $\{\alpha_m \}$ are chosen as the bases. 
We will adopt the standard terminology in group theory so that to avoid 
possible ambiguities. Meanwhile we will give possible identifications in 
terms of the standard terminology in quantum mechanics 
when we deal with any concrete Lie algebra.
We normalized the simple roots to unity so that the structure constant
in eq. (\ref{eq:ChavelleyBasis})
differs from the Cartan matrix in textbooks of group theory by a factor
1/2. 
 
{\it The cases of $B_2$ and $C_2$ Lie algebra}: 
As $C_2$ is isomorphic to $B_2$, we will only make our discussions on
$B_2$ Lie algebra. From the Dynkin diagram:
$$
\setlength{\unitlength}{1mm}
\begin{picture}(18,3.5)(-5,0)
 \linethickness{0.5pt}
  \put(0,1.1){\circle{2}}
   \put(1,0.6){\line(1,0){7.8}}\put(1,1.6){\line(1,0){7.8}}
    \put(10,1.1){\circle{2}}
\end{picture}
$$ 
which means that the $\alpha_1$ and $\alpha_2$ span an angle of $135^o$,
we can write out the simple roots in the nonorthogonal coordinates,
accordingly $\alpha_1=(1, -1/2)$, $\alpha_2=(-1, 1)$.
The set of roots for $B_2$ is
$\Delta=\{\pm\alpha_1, \pm\alpha_2, \pm(\alpha_1+\alpha_2),
\pm(2\alpha_1+\alpha_2)\}$,
The SO(5) Heisenberg chain is a chain with states on each site 
carrying a five dimensional representation of $B_2$ Lie algebra. These
five states ($|l\rangle, \, l=1,2,...,5$) 
are labeled by the eigenvalues of the 
Cartan subalgebra, which are two dimensional vectors called the 
weight vectors in group theory, namely,
$(0, 1/2)$, $(1,-1/2)$, $(0, 0)$, $(-1, 1/2)$, and $(0, -1/2)$.
For example $(H_1, H_2)|2\rangle=(1, -1/2)|2\rangle$.
If placing the states that carry the spinor representation of $B_2$
(meanwhile the fundamental repreaentation of $C_2$) Lie algegra
at each site, we will have a Sp(2) Heisenberg chain. It is a 
four dimensional representation labeled by weight vectors
$(1/2, 0)$, $(-1/2, 1/2)$, $(1/2, -1/2)$, and $(-1/2, 0)$ respectively.

Let us investigate the nature of the ground state of those systems.
We will extend the strategy of \cite{LSM} to show the ground state of 
present models on bipartite lattice $L=A\bigcup B$ is nondegenerate. 
By making use of a unitary transformation,
\begin{equation}
U=\exp\Bigl[-i\pi\sum_{y\in B}\bigl(H_1(y)+\frac{1}{2}\bigr)\Bigr],
\end{equation}
that rotates each state on the sublattice $B$, the original Hamiltonian 
(\ref{eq:Hamiltonian}) is mapped to the following form:
\begin{eqnarray}
\widetilde{\cal H}=\sum_{\stackrel{x,x'}{m n}}g^{mn}H_m(x)H_n(x')
 &-&\sum_{<x x'>}\biggl[\sum_{\tilde{\alpha}}
   E_{\tilde{\alpha}}(x) E_{-\tilde{\alpha}}(x')\nonumber\\
-E_\beta(x)E_{-\beta}(x')
 &-&E_{-\beta}(x) E_{\beta}(x')\biggr],
\label{eq:tildeHamiltonian}
\end{eqnarray}
where 
$\tilde{\alpha}\in\{\pm\alpha_1,\pm\alpha_2, \pm(2\alpha_1+\alpha_2)\}$
and $\beta=\alpha_1+\alpha_2$.
It is easy to show, after some algebra, that this transformation is 
also a canonical transformation.
The application of canonical transformation to the traditional
SU(2) Heisenberg model can turn out an overall negative sign in the second 
sum of eq. (\ref{eq:Hamiltonian}), but it is not possible in present case.
Nevertheless, we will see that this does not affect the proving of the
nondegenerate ground state although several authors had not succeeded.

As we consider $N=2n$ (bipartite lattice), the group theory 
concludes that there always exists one state of any multiplet lying in
the subspace of zero weight $(0,0)$, i.e., zero eigenvalues of 
$H_m^{tot}=\sum_x H_m (x)$, $m=1,2$.
This guarantees the eigenvalues determined within the subspace
cover the whole spectrum of the model.
A complete set of states in the subspace consists of all possible 
configurations that can be constructed in the following way.
As to Sp(2), there are $n_1$ sites labeled by 
$(1/2, 0)$ and the same number of sites labeled by $(-1/2, 0)$,
additionally $n_2$ sites labeled by 
$(-1/2, 1/2)$ and the same number of sites labeled by $(1/2, -1/2)$
for any partition $n=n_1+n_2$. 
For SO(5) however, we should consider arbitrary partitions 
$n=n_0+n_1+n_2$. The possible states in the subspace consists of 
$2n_0$ sites labeled by $(0,0)$,
$n_1$ labeled by $(0,1/2)$ and the same number of sites labeled by $(0,-1/2)$,
and $n_2$ sites labeled by $(1,-1/2)$ 
and the same number of sites labeled by $(-1,1/2)$. 

Denoting those states by $|\mu\rangle$, we can expand any eigenstate of 
$\widetilde{\cal H}$ in this subspace as 
$|\psi\rangle=\sum \langle\mu|\psi\rangle|\mu\rangle$. 
The Schr\"odinger equation 
$\widetilde{\cal H}|\psi\rangle=E|\psi\rangle$ in this representation 
gives rise to
\begin{equation}
\sum_{<x x'>}\eta(p_{x x'})\langle p_{x x'}(\mu)|\psi\rangle
 =(\epsilon_\mu-E)\langle\mu|\psi\rangle,
\label{eq:SchEq}
\end{equation}
where $\epsilon_\mu|\mu\rangle=\sum g^{mn}H_m(x)H_n(x')|\mu\rangle$ and 
$p_{xx'}$ stands for an exchange of the states on adjacent sites $x$ and $x'$. 
For Sp(2), $\eta(p_{xx'})=-1$ if the exchange occurs 
either between $(1/2,0)$ and $(1/2,-1/2)$, 
or between $(-1/2,1/2)$ and $(0,-1/2)$;
$\eta(p_{xx'})=1$ if for the other exchanges. 
For the convenience in the following discussions, 
we call the former the mutable exchange, 
the later the immutable exchange.
For SO(5) the mutable exchanges occur between
$(0, 1/2)$ and $(0, 0)$, or between  $(0, 0)$ and $(0, -1/2)$.

First we will show that all the coefficients 
$a_\mu=\langle\mu|\psi_0\rangle$ are nonvanishing for any ground 
state $|\psi_0\rangle=\sum a_\mu|\mu\rangle$. 
To prove this, we suppose some of them being zero, 
saying $a_{\bar{\mu}}=0$, and consider a trial state (wave function) 
$|\psi'\rangle=\sum \eta(\mu)\mid a_\mu\mid |\mu\rangle$ with 
$\eta(\mu)=\pm 1$.  
The $\eta(\mu)$ is defined in the following way. 
Given one state $|\mu_0\rangle$ for each afore-mentioned partition
in the subspace of null weight, any others of the whole states in 
the subspace can always be obtained by a sequence of adjacent permutations. 
We define $\eta(\mu)=1$ if 
even number of mutable exchange 
is involved in achieving the desired state $|\mu\rangle$. 
Otherwise, if odd number of that is involved, we define $\eta(\mu)=-1$. 
Now it is easy to calculate that
\begin{eqnarray}
\langle\psi'|\widetilde{\cal H}|\psi'\rangle
 =\sum_\mu a_\mu^2-\sum_\mu\sum_{<xx'>}
 \mid a_\mu\mid\mid a_{p_{xx'}(\mu)}\mid, \nonumber\\
\langle\psi_0|\widetilde{\cal H}|\psi_0\rangle
 =\sum_\mu a_\mu^2-\sum_\mu\sum_{<xx'>}
  \eta(p_{xx'})a_\mu a_{p_{xx'}(\mu)},
\label{eq:gsEnergy}
\end{eqnarray}
which concludes that 
\begin{equation}
\langle\psi'|\widetilde{\cal H}|\psi'\rangle
 \leq\langle\psi_0|\widetilde{\cal H}|\psi_0\rangle.
\label{eq:lessThan}
\end{equation}
Because $\mid a_{\bar{\mu}}\mid=0$ 
but $\sum \mid a_{p_{xx'}(\bar{\mu})}\mid\neq 0$,
$|\psi'\rangle$ is not an eigenstate.
According to variational principle we will have 
$\langle\psi'|\widetilde{\cal H}|\psi'\rangle > 
  \langle\psi_0|\widetilde{\cal H}|\psi_0\rangle$
that contrasts  with eq.(\ref{eq:lessThan}).
This contradiction proves  that
$a_{\bar{\mu}}=0$ is not possible for ground state. 

Clearly, if $|\psi_0\rangle$ is a ground state, only the equal sign in
eq. (\ref{eq:lessThan}) is possible. This implies that the coefficients
of two configurations/states should have opposite signs 
if they are related by odd number of mutable exchanges 
of adjacent sites, otherwise they should have the same sign.
It is obviously not possible to get two states 
with the mentioned restrictions on the coefficients 
to be orthogonal to each other. 
Therefore there can be only one ground state.      
Now we complete our prove that the ground state of Sp(2) and SO(5) 
Heisenberg models on bipartite lattice have nondegenerate ground state.

We are now in the position to observe the features of energy 
excitations above the nondegenerate ground state.
Introduce a slowly varying twist operator
\begin{equation}
V(\theta)=\exp\biggl[-i\theta\sum_{x=1}^{N}x K(x)\biggr],
\label{eq:twistB2}
\end{equation}
with $K(x)=H_1(x)$. In order to guarantee the periodic boundary condition,
we must let $\theta=2\pi\nu/N$ ($\nu$ is any integer number).
Since the ground state $|\psi_0\rangle$ is nondegenerate and the Hamiltonian
is invariant under translation, 
we have $T|\psi_0\rangle=e^{-i\delta}|\psi_0\rangle$ 
where $T$ denotes the operator of translation by one site.
Following Lieb, Schulz and Mattis \cite{LSM}, we construct a state
$|\psi_\nu\rangle=V(2\pi\nu/N)|\psi_0\rangle$. Noting the fact that
$\sum_x H_1(x)|\psi_0\rangle=0$, we get
\begin{eqnarray}
\langle\psi_0 | \psi_\nu\rangle&=&
  \langle\psi_0 | T V T^{-1}|\psi_0\rangle \nonumber\\
&=&\langle\psi_0 |V \exp\Bigl(i2\pi\nu K(1)\Bigr)|\psi_0\rangle.
\label{eq:orthogonal}
\end{eqnarray}
Obviously, 
$\langle\psi_0|\psi_\nu\rangle=-\langle\psi_0|\psi_\nu\rangle$
for an odd $\nu$ in the Sp(2) case, but 
$\langle\psi_0|\psi_\nu\rangle=\langle\psi_0|\psi_\nu\rangle$ 
for any $\nu$ in the SO(5) case.
So the state $|\psi_\nu\rangle$ of odd $\nu$
is orthogonal to the ground state $|\psi_0\rangle$ and hence is 
an excited state for the Sp(2) model, but is not for SO(5) one.

From the commutation relation eq. (\ref{eq:ChavelleyBasis})
for $B_2$ Lie algebra, we get,
\begin{eqnarray}
e^{i\theta H_1}E_{\pm\alpha_m}e^{-i\theta H_1}
  &=&e^{\mp i(-1)^m\theta} E_{\pm\alpha_m},\,\, m=1,2,\nonumber\\
e^{i\theta H_1}E_{\pm(\alpha_1+\alpha_2)}e^{-i\theta H_1}
  &=&E_{\pm(\alpha_1+\alpha_2)},\nonumber\\
e^{i\theta H_1}E_{\pm(2\alpha_1+\alpha_2)}e^{-i\theta H_1}
  &=&e^{\pm i\theta}E_{\pm(2\alpha_1+\alpha_2)}.\nonumber
\end{eqnarray}
With the help of these relations, we obtain after some algebra that
\begin{eqnarray}
V^\dagger{\cal H}V-{\cal H}=i\sin(2\pi\nu/N)
\Bigl[\sum_x xH_1(x),\; {\cal H}\Bigr]
  \nonumber\\
-2\sin^2(\pi\nu/N)\sum_{x}\sum_{\tilde{\alpha}}
  E_{\tilde{\alpha}}(x)E_{-\tilde{\alpha}}(x+1),
\label{eq:difference}
\end{eqnarray}
where 
$\tilde{\alpha}\in\{\pm\alpha_1,\pm\alpha_2, \pm(2\alpha_1+\alpha_2)\}$ and
the term corresponding to $\alpha_1+\alpha_2$ is absent in the summation.
Then the excitation energy is evaluated, 
\begin{equation}
\langle\psi_\nu|{\cal H}|\psi_\nu\rangle
 -\langle\psi_0|{\cal H}|\psi_0\rangle
  \leq\frac{2\pi^2}{N}\nu.
\label{eq:energyGap}
\end{equation}
Thus there is no energy gap in Sp(2) Heisenberg model.
For SO(5) however, 
the possibility of existence of energy gap could not be ruled out.

{\it The case of $A_3$ Lie algebra}:
From its Dynkin diagram 
$$
\setlength{\unitlength}{1mm}
\begin{picture}(18,3.8)(8,0)
 \linethickness{0.5pt}
 \put(2,2){\circle{2}}\put(3,2){\line(1,0){7.8}}
 \put(12,2){\circle{2}}\put(13,2){\line(1,0){7.8}}
 \put(22,2){\circle{2}}
\end{picture}
$$
we write out the simple roots in the nonorthogonal coordinates,
$\alpha_1=(1, -1/2, 0)$, $\alpha_2=(-1/2, 1, 1/2)$ and 
$\alpha_3=(0, -1/2, 1)$. 
In SU(4) Heisenberg model the states at each site carry out the 
fundamental representation of $A_3$ Lie algebra. 
The weight vectors of the four states are 
$(1/2, 0, 0)$, $(-1/2, 1/2, 0)$, $(0, -1/2, 1/2)$ and $(0, 0, -1/2)$.
In SO(6) model the six states at each site are labeled by weight vectors
$(0, 1/2, 0)$, $(1/2, -1/2, 1/2)$, $(-1/2, 0, 1/2)$,
$(1/2, 0, -1/2)$ $(-1/2, 1/2, -1/2)$ and $(0, -1/2, 0)$ respectively.
These states carry out the 6-dimensional representation of $A_3$ Lie algebra.

In ref.\cite{ALieb}, the assupmtion of unique ground state was made 
for investigating the excitations, here we can prove the nondegenerate
ground state rigorously.
Analogous to our previous discussion on the models
of $B_2$ Lie algebra, we consider again the model on bipartite lattice
and employ the following canonical transformation,
\begin{equation}
U=\exp\Bigl(-i\pi\sum_{y\in B}K(y)\Bigr),
\label{eq:A3canonical}
\end{equation}
here $K(y)=H_1(y)+H_3(y)$. This transformation maps the Hamiltonian into,
\begin{eqnarray}
\widetilde{\cal H}&=&\sum_{\stackrel{<x x'>}{m n}}g^{mn}H_m(x)H_n(x')
 -\sum_{<x x'>}\biggl[\sum_{\tilde{\alpha}}
   E_{\tilde{\alpha}}(x) E_{-\tilde{\alpha}}(x')
    \nonumber\\
\,&\, &\;\; -\sum_{\beta}E_{\beta}(x) E_{-\beta}(x')\biggr],
\label{eq:A3tildeH}
\end{eqnarray}
where $\widetilde{\alpha}\in\{\pm\alpha_1, \pm\alpha_2, 
\pm\alpha_3, \pm(\alpha_1 +\alpha_2 + \alpha_3)\}$ and 
$\beta\in\{\pm(\alpha_1 + \alpha_2), \pm(\alpha_2 + \alpha_3)\}$

Because we discuss the fundamental representation of $A_3$ Lie algebra
(instead of self-conjugate representation considered in \cite{ALieb}),
we need to consider $N=4n$. In this case, any multiplet of the 
system will always have a state within the subspace of zero
weight $(0, 0, 0)$. Any eigenstate of the $\widetilde{\cal H}$
in eq. (\ref{eq:A3tildeH}) can be expanded by $|\mu\rangle$ 
$\mu=1, 2,...,(4n)!/(n!)^4$. The Schr\"odinger equation in 
this representation is formally the same as eq. (\ref{eq:SchEq}).
The only difference is that the mutable exchanges  
occur between the states of either $(1/2, 0, 0)$ and $(0, -1/2, 1/2)$
or $(-1/2, 1/2, 0)$ and $(0, 0, -1/2)$. 
The formulations up to Eq. (\ref{eq:gsEnergy}) are almost the same 
as previous formulation except the $\eta(\mu)$ is defined according to
the mutable exchange of present $A_3$ representation.
Actually, we can also choose either $K=H_3-H_1$ or $K=H_1+2H_2+H_3$
(the set of $\tilde\alpha$ and $\beta$ variants correspondingly) 
to achieve the same conclusion that the ground states of SU(4) 
Heisenberg model and SO(6) one are nondegenerate.

Concerning the features of energy excitations above the nondegenerate
ground state, we should introduce slowly varying twist operator like
(\ref{eq:twistB2}) with $K=H_1 + H_3$.
Repeating the similar calculation we got formally the same relation like
Eq. (\ref{eq:orthogonal}).  Because the operator $H_1 + H_3$ acting on
all the SU(4) states will always yield eigenvalue of $1/2$, but acting on
all the SO(6) states will get $0$ or $1$. Then the state constructed
by the twist operator $V$ is orthogonal to the ground state in SU(4) case,
but is not in SO(6) case.
By means of the commutation relations (\ref{eq:ChavelleyBasis}) for 
$A_3$ Lie algebra, we obtain again Eq. (\ref{eq:energyGap}) after careful
calculation. We therefore conclude that there are gapless excitations 
above the nondegenerate ground state for SU(4) Heisenberg model
and suspect an energy gap opens up  in the SO(6) Heisenberg model.
It is worthwhile to point out that the above formulation can be 
extended to the fundamental representation for any SU(M) straitforwardly  
as long as the number of site is $N=nM$.

{\it The physics implications of those models}: Up to now, we adopted 
mathematical terminology so as to keep the discussions rigorous.
It is worthwhile to exhibit the physics implications of 
those models. We know the SU(4) Heisenberg model \cite{LiMSZ} describes
the spin system with twofold orbital degeneracy, which is an effective model
of doubly degenerate electrons at quarter-filling \cite{LiE} 
in the limit of strong on-site coupling. At half-filling, moreover, it 
reduces to a SO(6) Heisenberg model \cite{LiE} in the strong coupling limit.
The gapless nature of the SU(4) model was also confirmed 
in ref.\cite{LiMSZ} on the basis of Bethe-ansatz solution.

The Sp(2) and SO(5) Heisenberg models are discussed separately.
Consider the state on each site being double occupancy
of electrons, spin up, spin down and empty. It is not difficult to verify
that the four states 
$|1\rangle=|\uparrow\downarrow\rangle$, 
$|2\rangle=|\uparrow\rangle$, 
$|3\rangle=|\downarrow\rangle$ and 
$|4\rangle=|0\rangle$
carry out the fundamental representation of $C_2$ Lie algebra.
The Chevalley basis of the $C_2$ Lie algebra is realized by
$H_1=S^z$, $H_2=C^\dagger_\downarrow C_\downarrow-1/2$, 
$E_{\alpha_1}=S^+$ and $E_{\alpha_2}=C^\dagger_\downarrow$,
where $C^\dagger_\downarrow$ denotes the operator that creates
an electron of spin down. This give us a Sp(2) system.

The SO(5) system can be realized by pseudo-spin one particles. 
Excluding the double occupancy of parallel pseudo-spins, we can define 
$|1\rangle=|\Uparrow\Downarrow\rangle$, 
$|2\rangle=|\Uparrow\rangle$, 
$|3\rangle=|\Rightarrow\rangle$, 
$|4\rangle=|\Downarrow\rangle$ and
$|5\rangle=|0\rangle$
to carry out a 5-dimensional representation of $B_2$ Lie algebra.
The generators $H_1=J^z$, 
$H_2=C_\Rightarrow C^\dagger_\Rightarrow
 (C^\dagger_\Downarrow C_\Downarrow-1/2)$, 
$E_{\alpha_1}=J^+$ and 
$E_{\alpha_2}=C^\dagger_\Downarrow C_\Rightarrow C^\dagger_\Rightarrow$,
where $C^\dagger_\Downarrow$ and $C^\dagger_\Rightarrow$ 
create the state $|\Downarrow\rangle$ and $|\Rightarrow\rangle$ 
respectively so that  
$J^z|\Downarrow\rangle=-|\Downarrow\rangle$ and
$J^z|\Rightarrow\rangle=0$.
Here the creation/annihilation operators  are required to obey 
anti-commutation relations so as to realize commutation relations 
of the $B_2$ Lie algebra. 
Five states related to two $d$-wave superconducting order parameters
and three antiferromagnetic order parameters were suggested to constitute 
the bases of a SO(5) theory\cite{Zhang} for a phenomenological understanding
of the phase diagram of high $T_c$ superconductive materials. 

{\it In summary}, 
we extended the  Lieb-Schultz-Mattis theorem to a hierarchy of generalized
Heisenberg models related to non-exceptional Lie algebras.
The nondegenerate ground state in these models is rigorously proven
by means of a procedure analogous to the original LSM theorm 
for SU(2) model.
The main sketch of the proof consists of the following steps. 
Since the canonical transformation does not change the
spectrum of a system, as first step, we have found useful 
canonical transformations that map the original Hamiltonians 
into what can be 
analyzed by the method invented in \cite{LSM}.  
To confirm the nondegenerate ground state, we have proven that it is
impossible to construct second state which possesses the lowest
energy meanwhile keeps orthogonal to the given ground state.
As the standard literature employed by many authors 
\cite{LSM,ALieb,OYAffleck97a,OYAffleck97b}, we separately introduced
slowly varying twist operators for those models.
The twist operator will create a gapless excitation mode from the 
nondegenerate ground state, as long as the created state is orthogonal
to the original ground state. Otherwise, the possibility of existence of
an energy gap can not be ruled out. 
Our investigation have shown that there is no energy gap above the 
ground sates in SO(4), Sp(2) and SU(4) Heisenberg models, 
but gaps are suspected to open in SO(5) and SO(6) models. 

The work is supported by NSFC No.1-9975040 and EYF of China Ministry of
Education, also supported by AvH Stiftung.
The author thanks  M.Ma for beneficial discussions at the early stage 
of present work; also thanks I.Affleck, F.D.M.Haldane, 
M.Oshikawa, and F.C.Zhang for interesting discussions.

\end{multicols}


\begin{thebibliography}{99}

\bibitem{Marshall}
W. Marshal, Proc. Roy. Soc. London A {\bf 232}, 48 (1955).

\bibitem{LSM}
E.H. Lieb, T.D. Schulz, and D.C. Mattis, 
Ann. Phys. (N.Y.) {\bf 16}, 407 (1961).

\bibitem{Haldane}
F.D.M. Haldane, Phys. Lett. A {\bf 93}, 464 (1983);
Phys. Rev. Lett. {\bf 50}, 1153 (1983).

\bibitem{Kolb}
M. Kolb, Phys. Rev. B {\bf 31}, 7494 (1985).

\bibitem{ALieb}
I. Affleck and E.H. Lieb, Lett. Math. Phys. {\bf 12}, 57 (1986).

\bibitem{OYAffleck97a}
M. Oshikawa, M. Yamanaka, and I. Affleck, Phys. Rev. Lett. {\bf 78}, 
1984 (1997).

\bibitem{OYAffleck97b}
M. Oshikawa, M. Yamanaka, and I. Affleck, Phys. Rev. Lett. {\bf 79}, 
1110 (1997).
M. Oshikawa, Phys. Rev. Lett. {\bf 84}, 1535 (2000).

\bibitem{LiMSZ}
Y.Q. Li, M. Ma, D.N. Shi, and F.C. Zhang,
Phys. Rev. Lett. {\bf 81}, 3527(1998); 
Phys. Rev. {\bf B 60}, 12781(1999).

\bibitem{Zhang}
S.C. Zhang, Science {\bf 275}, 1089 (1997).
 
\bibitem{TNagaosa}
Y. Tokura and  N. Nagaosa, Science 288, 462 (2000). 

\bibitem{Gilmore} {\it e.g.}
R. Gilmore, 
{\it Lie groups, Lie algebras and some of their applications},
John Wiley \& Sons, New York, 1974.

\bibitem{LiE}
Y.Q. Li and U. Eckern, Phys. Rev. B 62, 15493 (2000).

\end{thebibliography}
\end{document}